\documentclass{article}

\usepackage{arxiv}

\usepackage[utf8]{inputenc} 
\usepackage[T1]{fontenc}    
\usepackage{hyperref}       
\usepackage{url}            
\usepackage{booktabs}       
\usepackage{amsfonts}       
\usepackage{nicefrac}       
\usepackage{microtype}      
\usepackage[utf8]{inputenc}
\usepackage{graphicx}
\usepackage{natbib}
\usepackage{doi}
\usepackage{authblk}
\usepackage{subcaption}
\usepackage{multirow}
\usepackage{bm}
\usepackage{siunitx}
\usepackage{soul}
\usepackage{amsmath}

\usepackage{color}

\usepackage{amssymb}
\newcommand*{\circledbullet}{%
    \mathbin{%
        \ooalign{$\circledcirc$\cr\hidewidth$\bullet$\hidewidth}%
    }%
}

\renewcommand{\shorttitle}{Seamless prediction of high-impact weather events}
\title{Seamless prediction of high-impact weather events:\\ a comparison of actionable forecasts}
\author{Zied Ben Bouall\`egue}

\affil{European Centre for Medium-Range Weather Forecasts}

\begin{document}
\maketitle

\begin{abstract}

A new index for high-impact weather forecasting is introduced and assessed in comparison with the well-established extreme forecast index (EFI).
Two other ensemble summary statistics are also included in this comparison study: the shift-of-tail and a standardised ensemble mean anomaly. 
All these forecasts are based on the same ingredients: 
the ensemble forecast run at the European Centre for Medium-Range Weather Forecasts 
and the corresponding model climatology derived from a set of reforecasts.
The new index emerges from recent developments in forecast verification of extreme events: it is derived as 
a consistent forecast with the diagonal score, a weighted version of the continuous ranked probability score targetting high-impact events. 
In this study, we emphasise the importance of forecast discretisation for communication purposes and decision-making. A forecast is actionable  
in the situation where a user can decide to take action when a threshold is exceeded by the forecast. 
Forecast verification is performed to assess both the potential skill of the different indices as well as their specific skill as actionable forecasts.
Among the investigated actionable forecasts, the new proposed index demonstrates the strongest discrimination power, in particular at longer lead times,
paving the way for seamless predictions of high-impact weather across time ranges.

\end{abstract}

\section{Introduction}
\vspace*{-1pt}
\noindent

Ensemble weather forecasts consist of a set of weather scenarios that capture a range of equally probable (or at least possible) outcomes. 
Ensembles serve as a basis for quantifying forecast uncertainty and generating probabilistic forecasts. 
Our ability to predict high-impact weather events using ensemble forecasts has improved over the past decades \citep{zbb19}.
However, the use and communication of probabilistic forecasts for effective decision-making is still in its infancy, 
in particular when it comes to ``low-probability, high-impact'' events \citep{fundel2019}.

At the beginning of the century, the extreme forecast index (EFI) was introduced by \cite{lalaurette2003} to provide support for generating early warnings of extreme events.
This index highlights situations related to ``unusual'' forecasts by comparing an ensemble forecast distribution with a model climate distribution.
The EFI is an indicator of the likelihood of an extreme event but doesn't provide information about the intensity of the event.
For this reason, the EFI is often complemented by the shift-of-tail (SOT), a normalised quantile forecast
using the climatological quantiles for normalisation \citep{zsoter2006}.    

Several studies have illustrated that  EFI has predictive skill not only for
raising early awareness of extreme precipitation \citep[][in particular in case of severe convection, \citealt{tsonevsky2018}]{lavers2016,lavers2017,lavers2018}, but also for 
the early warning of extreme winds and extreme windstorms \citep{petroliagis2014,boisserie2016}.
SOT predictive performance is less explored in the literature but seems to show a comparable level of skill as the EFI \citep{boisserie2016,raynaud2018}.  
Another ensemble summary statistic has also been suggested for the early detection of extreme weather: the ensemble mean anomaly forecast \cite[ANF,][]{guan2017}.  
ANF appears highly correlated to the EFI for temperature with slightly higher skill in predicting extreme cold events.

In this study, we introduced and assessed a new index for high-impact weather forecasting. 
We are building on the concept of crossing-point forecast (CPF) that has emerged from recent developments in forecast verification of extreme events \citep{zbb2021}.
A CPF is a consistent forecast with the diagonal score which is a weighted version of the continuous ranked probability score \citep{zbb2017}.
The diagonal score is routinely used at ECMWF for assessing the performance of the ensemble forecasting system \citep{haiden2021}, 
but the underlying actionable forecast has not been exploited so far.

An actionable forecast is defined as a forecast discretised with a set of decision thresholds such that
a user can decide to take action when a threshold is exceeded by the forecast. Seamless forecasting across lead times would rely 
on actionable forecasts that could be used across time ranges, without the need to change the forecast definition or the set of decision thresholds 
with the forecast lead time. In the following, the focus is on actionable forecasts derived from an ensemble prediction system.

The CPF, like the EFI, is a summary statistic of both the ensemble forecast and its model climatology. 
In the CPF case, the focus is set on the intersection point between the two cumulative probability distributions.
This intersection marks the limit between the situation where the risk of an event is higher \textit{in general} 
(in the climatology, with no specific information about the current weather situation)
 or \textit{in particular} for a given day (in the forecast, given the information available at the start of the prediction).  
The CPF is a probabilistic forecast which takes value between 0 and 1 so one can use it as a forecast index. 
Here, we explore the CPF performance as a forecast index for extreme daily precipitation events.

This paper is organised as follows: 
a definition of the different types of forecasts compared in this study is provided in Section \ref{sec:data}, 
two case studies are shown and commented on in Section \ref{sec:examples}, 
forecast intrinsic characteristics and their interpretation are discussed in Section \ref{sec:chara},
general verification results are presented in Section \ref{sec:results} followed by a focus on actionable forecasts 
for seamless prediction of extreme events in Section \ref{sec:discuss} 
before to conclude in Section \ref{sec:conc}.

\section{From an ensemble forecast to actionable forecasts}
\label{sec:data} 


For high-impact weather forecasting, a set of ensemble-based indicators can be generated by comparing an ensemble forecast with a local model climatology.
A link between forecast intensity and potential impact is made by interpreting a forecast in the context of a local climate:
the climatological frequency of an event serves as a measure of its rareness and its potential impact. 
Also, the local variability of the weather can serve to contextualise the uncertainty of a forecast (as measured by the ensemble spread, for example).
Ensemble-derived indicators become actionable forecasts when discretised with a set of decision thresholds for visualisation and decision-making purposes 
as discussed in Section~\ref{sec:discuss}.

In the following, forecast and climatology take the form of cumulative probability distributions denoted $F$ and $G$, respectively. 
The forecast empirical distribution is derived from the 50-member ensemble forecast run operationally at ECMWF while the climate empirical distribution 
is derived from reforecasts which consist of 10 members run twice weekly over the past 20 years \citep{lalaurette2003}.
A local climatology is built using all reforecasts available within a time window of $\pm$2 weeks with respect to the validity date of the forecast.

The climate distribution that reflects the model climatology is referred to as the M-climate in opposition to the observed 
climate that is estimated based on observations (and used for verification).  In any case, \textit{climatology} is meant 
here in a local sense both in space and time: a climatology distribution is associated with a given grid point (or point observation) and a given day of the year.  
In practice, the empirical climate distributions are described by percentiles from 0\% (minimum) to 100\% (maximum) with 1\% intervals.

In this study, we compare 4 different types of forecasts, all derived from both $F$ and $G$.
CPF is based on the intersection point between $F$ and $G$, EFI is an integral of $F$ with respect to $G$, 
SOT focuses on one percentile of the tail of $F$ compared with the tail of $G$, while ANF is a standardised mean of $F$ using the first and second moments of $G$. 
We provide below a formal definition of each of these ensemble-derived forecasts.
 
\subsection{Crossing-point forecast}
\label{sec:cpf}
Consider the \textit{single crossing condition}:  $F$ and $G$ satisfy the single crossing condition if there exists a $y^\star$such that
\begin{equation}
\forall x, x \geq y^\star \Longrightarrow F(x) \geq G(x)
\end{equation}
and
\begin{equation}
\forall x, x \leq y^\star \Longrightarrow F(x) \leq G(x).
\end{equation}
When $y^\star$ exists, the crossing-point forecast is defined as:
\begin{equation}
\text{CPF} := G^{-1}(y^{\star}) 
\end{equation}
and is interpreted as a \textit{probabilistic worst-case scenario} \citep{zbb2021}.
CPF corresponds to the most extreme forecast event such that the risk in the forecast is larger than the climate risk. 
This event is expressed in terms of a quantile of the climate distribution. This quantile $q$ can easily be translated in terms of a local return period $r$. 
For example, if $q=0.95$, $r=20$ years, if $q=0.99$, $r=100$ years, and so on.

\begin{figure*}
\centering    
\includegraphics[width=0.4\textwidth,trim={0 0 -1cm 0},clip]{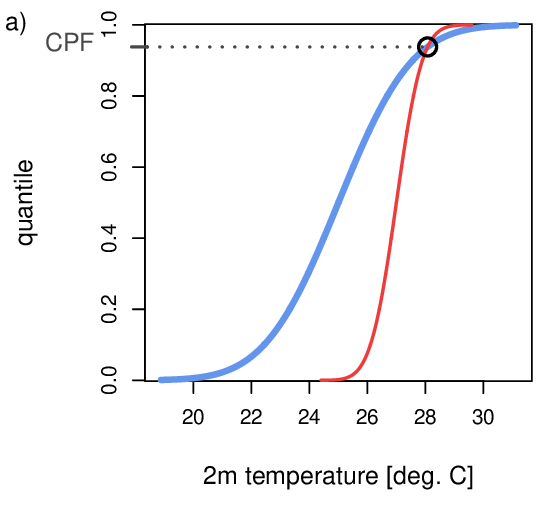}
\includegraphics[width=0.4\textwidth,trim={-1cm 0 0 0},clip]{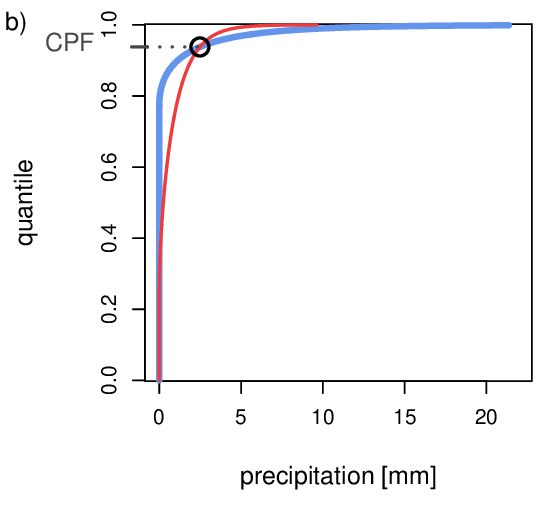}
\caption{
   Synthetic examples of CPF in the case of (a) temperature-like distributions and (b) precipitation-like distributions with $F$ in red and $G$ in blue.
}
\label{fig:cpf}
\end{figure*}

We use synthetic data to illustrate the concept of CPF in Fig.~\ref{fig:cpf}. 
While the plot on the left represents a canonical example, the plot on the right illustrates the difficulty of dealing with bounded variables. 
For precipitation (or wind), forecast and climate cdfs overlap at the point-mass 0. In that case, in practice, CPF is derived by scanning $F(x)$ and $G(x)$ with increasing $x>0$ (from left to right) 
until an intersection point is found (such that $F$ crosses $G$ from F<G to F>G, and not the other way around). 
When no intersection exists, CPF takes the value 0 or 1 depending on whether the $F$ is always above or below $G$, respectively.

\subsection{Extreme forecast index}
\label{sec:efi}

In \cite{zsoter2006} and \cite{tsonevsky2018}, the ensemble forecast index EFI is defined  as:
\begin{equation}
\text{EFI} := \frac{2}{\pi}\int_{0}^{1} \frac{p-F(p)}{\sqrt{p(1-p)}}\text{d}p 
\end{equation}
where $F(p)$ is the proportion of the ensemble members lying below the $p^{th}$-percentile of the climate distribution. The denominator helps provide more weight to the tails of the distribution. The index takes values in [-1,1]. The closer the EFI values to -1 or 1, the more abnormal the ensemble forecast and the more likely an extreme event.

\subsection{Shift of tails}
\label{sec:sot}
The shift of tails (SOT) is designed to compare the tails of $F$ and $G$. For the upper tail, the SOT is defined as:
\begin{equation}
\text{SOT} := -\frac{G^{-1}(0.99) - F^{-1}(0.9)} {G^{-1}(0.99) - G^{-1}(0.9)} 
\end{equation}
with $G^{-1}(\tau)$ and $F^{-1}(\tau)$ the $\tau$\%-quantile of the climate and forecast distribution, respectively. 
When SOT is greater than 0, it means that at least 10\% of the ensemble members are beyond the 99$^\text{th}$ M-climate percentile. 
The larger the SOT, the further these 10\% are from the M-climate. 
SOT was designed as a complementary forecast to the EFI, conveying a piece of information about the intensity of the expected extreme event.

\subsection{Anomaly forecast}
\label{sec:anc}
Anomaly forecasting focuses on the departure of the ensemble mean with respect to the climatology \citep{guan2017}.
The anomaly forecast (ANF) is here defined as the standardised ensemble mean anomaly with respect to the M-climate mean. The standardisation is based on
 the standard deviation of the M-climate distribution. ANF is computed as follows:
\begin{equation}
\text{ANF} := \frac{\mu(F) - \mu(G)} {\sigma(G) + k} 
\end{equation}
where $\mu$ and $\sigma$ are the mean and standard deviation functions, respectively, and $k$ is a constant set to 1 for precipitation 
(to avoid numerical issues with ANF computation for arid areas), 0 for temperature for example.

\section{Case studies}
\label{sec:examples}
Let's illustrate similarities and specificities between these difference forecasts with 2 case studies.
We compare forecasts  available 6 days and 1 day ahead of the event (lead time day 6 and day 1, respectively) but valid for the same dates.
A subjective assessment of the forecast performance at day 6 can be achieved by using the forecast at day 1 as a proxy of the \textquoteleft truth\textquoteright. 
The 2 case studies are related to high-impact events that have affected Europe over Summer 2021.

\subsection{Example 1: heavy precipitation over Belgium and Germany in July 2021}
\label{sec:extp1}
\begin{figure*}                                                        
\begin{center}                                                        
\includegraphics[width=0.95\textwidth]{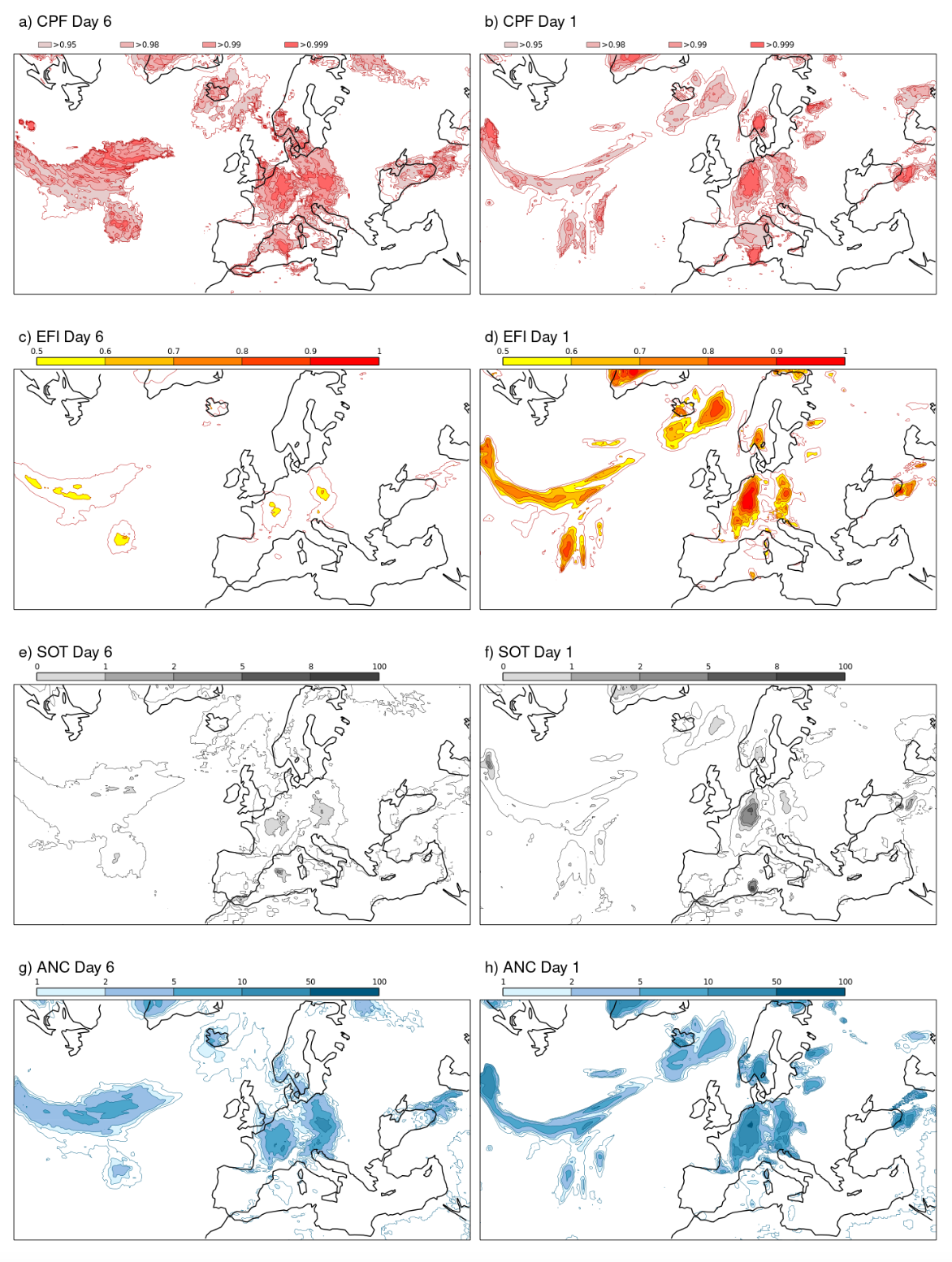}
\end{center}  
\caption{
	Daily precipitation forecasts valid on July 14, 2021:  (a,b) CPF, (c,d)  EFI, (e,f) SOT, and (g,h) ANC generated  6 days ahead of the event (left) 
    and 1 day ahead of the event (right).
}
\label{fig:flood}
\end{figure*}

Fig. \ref{fig:flood} shows CFP, EFI, SOT, and ANC of daily precipitation valid on July 14, 2021. 
Extreme rainfall on that day led to flash floods and large-scale flooding that had a catastrophic impact on parts of Belgium and western Germany. 

Forecasts 6 days ahead of the event are shown in the left panels: 
CPF has a large-scale signal with high values over the Atlantic and Western/Central Europe in Fig.~\ref{fig:flood}(a)
while EFI has a weak signal (if any) over the same areas in Fig.~\ref{fig:flood}(c).  Forecasts one day before the event are presented in the right panels.  
We see a convergence of CPF and EFI towards a similar solution with the strongest signal over the impacted areas in Figs~\ref{fig:flood}(b) and \ref{fig:flood}(d).

In this first example, the following general pattern emerges when comparing CPF and EFI:
\begin{enumerate}
\item CPF displays a strong signal at a longer lead time but the signal is large-scale and noisy,
\item EFI presents a weak signal or no signal at all at longer lead times,
\item both forecasts converge at shorter lead times.
\end{enumerate}

Regarding SOT and ANC, they have similar characteristics as EFI. Their signal appears smooth and/or weak 
at longer lead times in Figs~\ref{fig:flood}(e) and \ref{fig:flood}(g), respectively,
while a larger and sharper signal emerges closer to the event in Figs~\ref{fig:flood}(f) and \ref{fig:flood} (h). 

\subsection{Example 2: Convective precipitation over London in July 2021}
\label{sec:extp2}
\begin{figure*}
\begin{center} 
\includegraphics[width=0.95\textwidth]{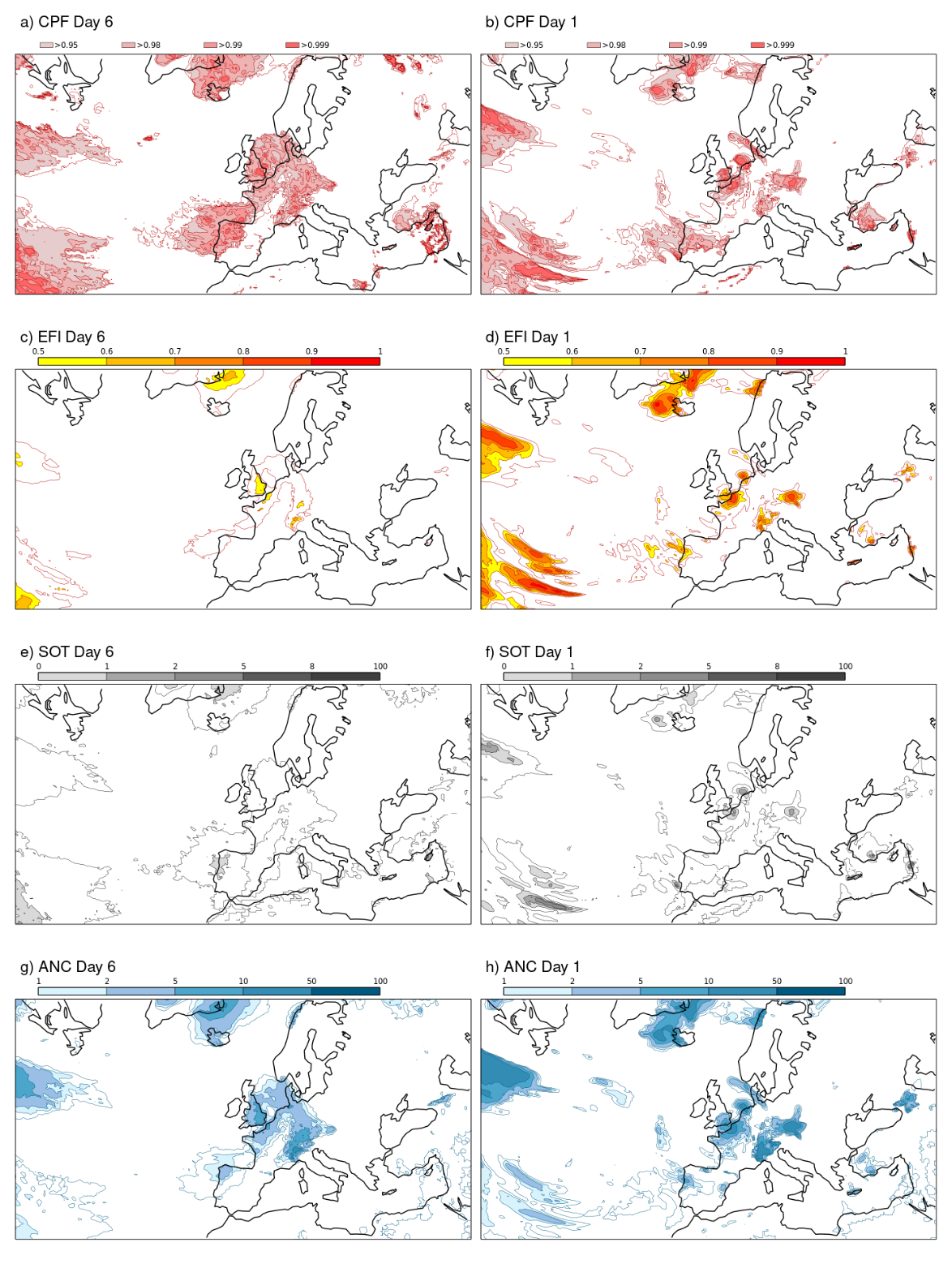}
\end{center} 
\caption{ 
    Same as Fig. \ref{fig:flood} but for forecasts valid on July 25, 2021. 
}
\label{fig:london}
\end{figure*}

Fig.~\ref{fig:london} compares actionable forecasts for precipitation on July 25, 2021. On that date, 
flash floods were reported in the London area after intense precipitation in South-East England. With respect to the previous example, 
this example deals with a more localised event, driven by convective activity.

Similarly to the previous example, we observe the following development with the forecast lead time as we approach the validity time: 
a noisy and large-scale CPF signal becomes more accurate while a smooth and light EFI signal evolves into a stronger and more focused signal. 
As shown in Figs~\ref{fig:london}(a) and \ref{fig:london}(c), the main differences between CPF and EFI are at longer lead times when the uncertainty in the ensemble forecast 
is large and the ensemble mean has not significantly departed from the climatological mean.

In this example, we can also notice differences between CPF and EFI at day 1, in particular over England, by comparing Figs~\ref{fig:london}(b) and \ref{fig:london}(d). 
Due to the convective nature of the intense precipitation over South-East England on that day, the predictability of the event was quite low. 
This low level of predictability was appropriately captured by the ensemble: the ensemble spread was large (not shown) with only some members predicting intense 
rain. As the uncertainty remains high even at a shorter lead time, the EFI never reached high values (greater than 50\%) even at day 1. 
The CPF signal over England was strong already at day 6 but much more refined over the impacted area at day 1.

\section{Forecasts intrinsic characteristics and interpretation}
\label{sec:chara}

\subsection{Correlation between forecasts}
\label{sec:quant}

\begin{figure*}
\begin{center}
\includegraphics[width=0.49\textwidth,trim={0 0 0 0.0cm},clip]{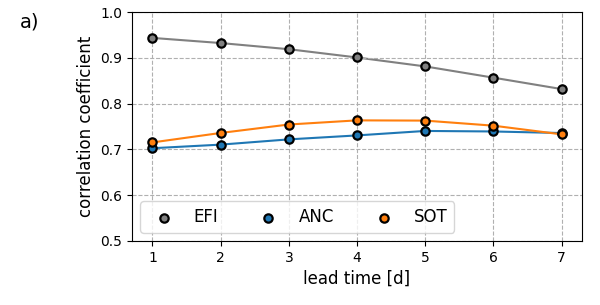}
\includegraphics[width=0.49\textwidth,trim={0 0 0 0.0cm},clip]{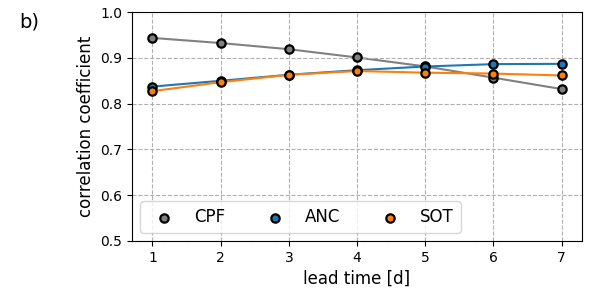}
\end{center}
\caption{
    Averaged rank correlation coefficients between daily precipitation forecasts over Europe during Summer 2021. 
    Correlation of (a) CPF and (b) EFI with respect to one another, SOT, and ANF. 
    The grey lines showing the correlation between CPF and EFI are identical in both panels. 
}
\label{fig:corco}
\end{figure*}

Following the visual inspection and comparison of CPF, EFI, SOT, and ANC in Figs \ref{fig:flood} and \ref{fig:london}, 
the forecasts are now compared quantitatively in terms of correlation. 
The level of similarity between the different forecasts is measured 
with the rank correlation coefficient (also known as Kendall-$\tau$ correlation coefficient). The results are shown in Fig. \ref{fig:corco}.

The correlation between CPF and EFI is 0.94 at day 1 but decreases almost linearly with the forecast lead time to reach 0.86 at day 6. 
Also, the correlation of CPF with ANF and SOT as shown in Fig. \ref{fig:corco}(a) is much smaller than the correlation of EFI with respect 
to the same quantities as shown in Fig. \ref{fig:corco}(b). These results suggest that the signal captured by CPF has different characteristics than 
the one exhibited by EFI, SOT, and ANF.

The correlation of EFI with ANF is close to 0.9 for all lead times. Because ANF is independent of the ensemble spread, this result confirms 
that EFI is mainly driven by changes in the ensemble mean with occasional modulations by the ensemble spread.
 EFI sensitivity to changes in the ensemble mean and the ensemble spread is discussed more thoroughly in \cite{dutra2013}. 
 The correlation of EFI with SOT is also high and slightly increases with lead time. SOT and ANF are themselves highly correlated (not shown) 
 which can be explained by the heteroscedastic nature of the precipitation distributions: variations in the ensemble mean and in quantiles tend to be strongly linked.

\subsection{Forecast distribution and forecast lead time}
\label{sec:evol}

\begin{figure*}
\begin{center}
\includegraphics[width=0.96\textwidth,trim={0 0 0 0.0cm},clip]{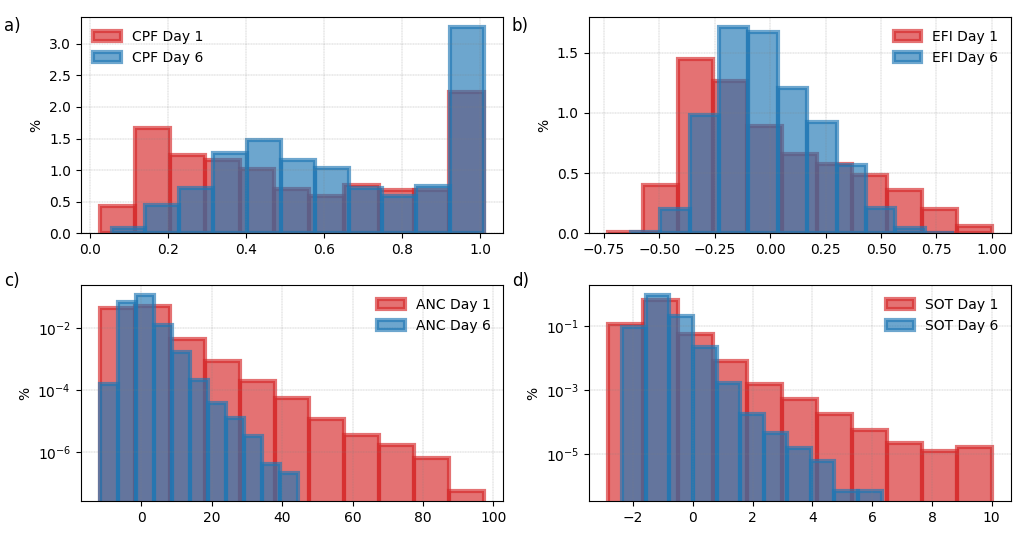}
\end{center}
\caption{
    Forecast distribution at day 1 (red) and day 6 (blue) over Europe for Summer 2021: (a) CPF, (b) EFI, (c) SOT and (d) ANC of daily precipitation. 
    Note the logarithmic scale of the \textit{y}-axis in (c) and (d). CPF is the only type of forecasts exhibiting stronger values at longer lead times. 
}
\label{fig:distri}
\end{figure*}

One key difference between the 4 types of forecasts investigated here is their evolution with forecast lead times
as illustrated in the 2 case studies in  Figs~\ref{fig:flood} and \ref{fig:london}.
To support a discussion on this point, we now analyse the forecasts' overall distributions during Summer 2021 for the European domain.
Fig.~\ref{fig:distri} compares the distributions at day 1 and day 6 for each forecast type.

CPF exhibits, in general, larger values at longer lead times as shown in Fig.~\ref{fig:distri}(a). 
Indeed, the category closer to 1 is the most populated at day 6. At day 1, the distribution of CPF values tends to resemble a uniform distribution:
the strong signal at longer lead times in CPF are tempered as we approach to the forecast validity time. 

In contrast, large values of EFI, SOT, and ANC are issued predominantly at shorter lead times as shown in Figs~\ref{fig:distri}(b,c,d), respectively.
The forecast uncertainty has a smoothing effect on these high-impact weather indicators such that large values 
are enabled only when the ensemble spread is small. This characteristic is a limiting factor to communicating effectively
low probability, high-impact events as discussed in Section~\ref{sec:discuss}.

\subsection{Forecasts interpretation}

CPF and EFI are both bounded forecasts taking values in [0,1] and [-1,1] respectively but they follow different interpretations.
EFI is an index, a number between -1 and 1 with no known statistical meaning attached to a single EFI value. 
Its interpretation is the following: the larger the EFI values the higher the risk of a high-impact event. 
The absence of a stringent statistical interpretation is an asset for use by a wider audience. 
However, one important limitation of the EFI is that no information is conveyed with this index about the event itself, its intensity, or its uncertainty
as already pointed out in previous studies \citep{neal2014,boisserie2016}.

In contrast, the CPF can be interpreted as an index but has also a meaning in statistical terms: it is a quantile level of a climate distribution.
Therefore, CPF values are directly related to local return periods (while EFI is somehow indirectly related to local return periods). 
For example, a CPF value of 0.95 corresponds to an event with a return period of 20 years with respect to the local climate. 
By expressing the forecast in return periods (or quantile level), systematic model errors are also automatically corrected while bypassing the need 
for a post-processing step involving observations \citep{fundel2010,prates2011}. 
In other words, a 20-year return event derived from a model climatology can be interpreted by the users based on an observed climatology.

We should also emphasise that a finer discretisation and a more accurate representation of the distribution tails 
would be necessary for a focus on \textit{very} extreme events.
So far, we have used a climatology described with  1\% quantile intervals and a forecast empirical distribution based on 50 ensemble members.
The relatively small ensemble size hinders an accurate estimation of extreme quantiles of the forecast distribution \citep{leutbecher2018,tempest2023}.
A finer discretisation of the tails could be achieved with the help of statistical methods based on extreme value theory 
and a more accurate representation of the forecast distribution tail by increasing the number of ensemble members
by leveraging, for example, the potential to generate large ensembles 
with data-driven models based on machine learning \citep{zbb2023bams}. 

\section{General performance characteristics}
\label{sec:results}


Assessing forecast performance for high-impact events is challenging because, by definition, these events are extreme and rare\footnote{Exposure
 and vulnerability are not considered in this study.}. Here, we consider \textit{relative} rather than \textit{absolute} extreme events: we focus on events with a \textit{local} return period of 20 years 
for a given point in space and time. In other words,  we focus on threshold-exceedance events where a threshold is defined as the 95$^\text{th}$ percentile of the local climate distribution. 
Following this approach and using synoptic observations as a reference, we mimic the standard EFI verification setup 
used to assess and communicate EFI performance as an ECMWF headline score \citep{tsonevsky2018,haiden2021}.
Here, the verification sample covers Europe over Summer 2021. 

The forecasts that we compare here differ by their very own nature. For example, CPF is a summary statistic consistent with a proper score while EFI is an index with no statistical meaning.
Aiming at a fair comparison, CPF and EFI are both considered actionable forecasts in the form of an index for the prediction 
of high-impact weather events. Both indices follow the same interpretation: a larger CPF (or EFI) value indicates a higher risk of a high-impact event materializing. 
Similarly, for SOT and ANC, a larger value in the forecast is interpreted as a higher risk of a precipitation event without directly taking into account the actual numerical value of the forecast.

Forecast performance is explored using contingency tables. A decision threshold is applied to the forecast to transform it into a binary (yes/no) forecast
while an observation is transformed into binary observation using a climate quantile as an event-threshold. 
For the discretisation of the forecast, a set of decision thresholds needs to be defined for each forecast type.
It follows that contingency tables are populated for each decision threshold of each forecast type. Eventually, 
hit rate ($H$) and false alarm rate ($F$) are estimated from these contingency tables to derive diagnostic plots and summary verification metrics. 
In particular, Relative Operating Characteristics (ROC) curves and economic value plots are generated and commented on below.

\subsection{Potential discrimination ability}
\label{sec:tp}

Discrimination is assessed within the ROC framework.  
A ROC curve is a set of points ($F$,$H$) for increasing decision-thresholds \citep[see][and references within]{zbb2022}.
Potential discrimination is assessed by using a large number of decision thresholds in order to span all possible forecast values\footnote{Alternatively, 
one could also examine distribution-fitted ROC curves \citep[see for example][]{zbb2022}, but this approach is not explored here.}.
Here we use 500 different decision thresholds equally spaced between the minimum and the maximum possible value of each forecast type. 
We talk about ``potential'' discrimination because, in practice, 500 different forecast values could not be distinguished on a map for decision-making.

\begin{figure*}
    \begin{center}   
    \includegraphics[width=0.95\textwidth]{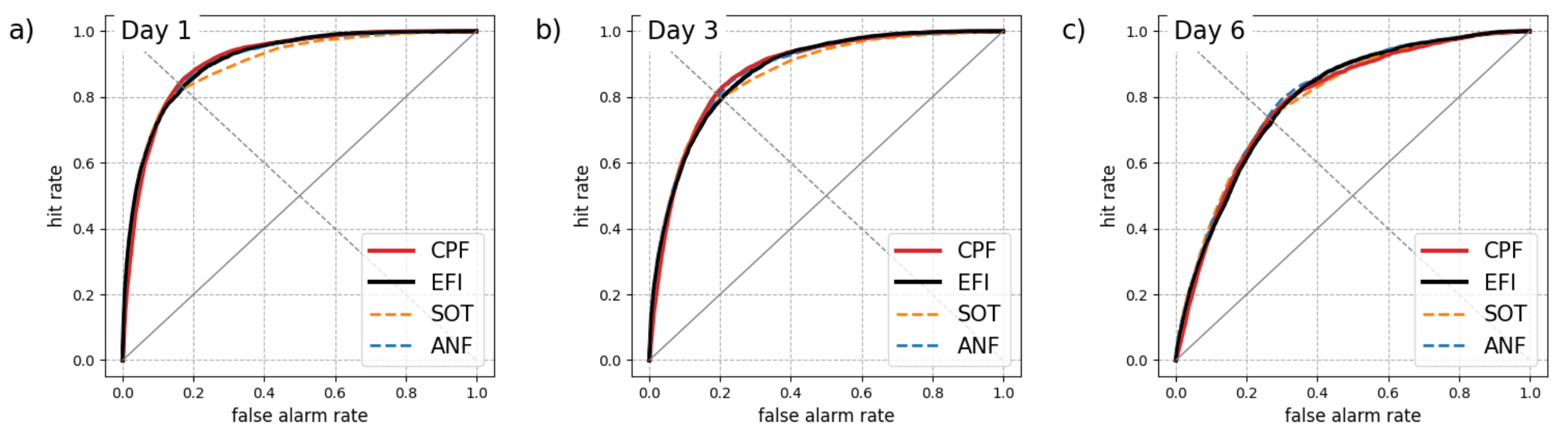}
    \end{center}
    \caption{ 
    ROC curves assessing the potential discrimination ability of CPF, EFI, ANF, and SOT for daily precipitation exceeding the 95\% climate percentile.  
    Resulst for forecast lead times (a) day1 , (b) day 3, and (c)  day 6.
    }
    \label{fig:pdf}
\end{figure*}

ROC curves based on 500 decision thresholds are shown in Fig.~\ref{fig:pdf} for CPF, EFI, SOT, and ANF at 3 different lead times.
The ROC curves are similar for all 4 types of forecasts indicating similar levels of potential discrimination ability, noting that 
CPF is slightly better and SOT slightly worse at the shorter lead time. Recalling the correlation coefficients in Fig.~\ref{fig:corco}, 
one could infer some level of complementarity between the different types of forecasts that could be further exploited in a post-processing framework. 

\subsection{Discriminating between moderate and heavy precipitation events }
\label{sec:value}

We propose now to compare the potential ability of the 4 different types of forecasts to distinguish between ``moderate'' and ``heavy'' precipitation events.  
We consider the situation where ``wet'' areas (with observed precipitation) can be identified in advance and assess 
the forecast's ability to pinpoint subareas with heavy precipitation.

In a ROC framework, the forecast is conditioned on the observation. For example, $H$, the hit rate, is an estimate 
of the probability of a correct forecast given that the event actually materialised. From a theoretical point of view,
discrimination is assessed based on a likelihood/base-rate factorization of the forecast/observation joint probability distribution \citep{murphy87}.
For this reason, the verification dataset can be stratified as a function of the observations while preserving the interpretability of the results\footnote{
Note that a stratification as a function of the observation only would violate the necessary condition for proper score estimations.}.

Here, we assess whether the forecast is able to discriminate between extreme precipitation events (observations exceeding the 95\% percentile) among lighter precipitation events
(observation exceeding the 70\% percentile of the local climate). For this purpose,
we build contingency tables from a verification dataset that contains only cases where the observation exceeds the 70\% percentile of the climate distribution.

\begin{figure*}
\begin{center}
  \includegraphics[width=0.95\textwidth]{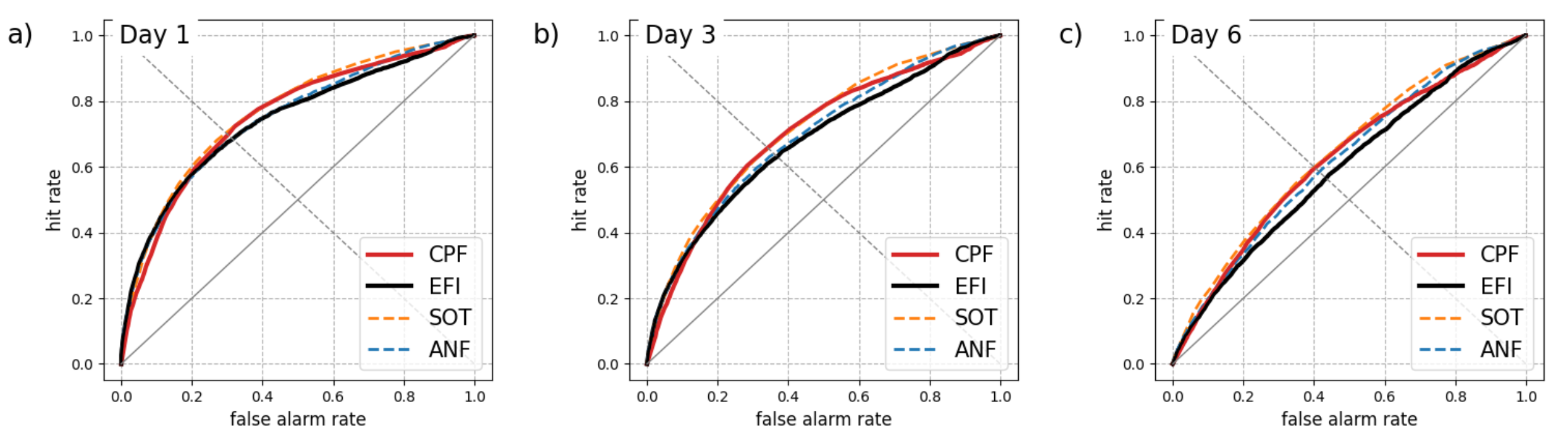}
\end{center}
\caption{ 
    Potential discrimination ability to distinguish between ``moderate'' and ``heavy'' precipitation events.
    Same as Fig. \ref{fig:pdf} but for the discrimination among the observed cases where precipitation exceed the 70\% climate percentile.  
}   
\label{fig:condi}
\end{figure*}

The results of this conditional verification exercise are shown in Fig.~\ref{fig:condi}. 
CPF and SOT appear better than ANF and EFI in this verification setting particularly at lead times day 3 and day 6. 
EFI is the type of forecast showing the smallest ability to discriminate between extreme and less extreme precipitation events, 
at these time ranges. 

\subsection{Potential economic value}
\label{sec:value}

The economic value of a forecast is estimated with the help of a standard cost-loss model. 
We consider the case where a forecast user can mitigate the loss associated with the occurrence of a weather event by taking preventive action.
The user decides to take action or not based on an actionable forecast. The key parameters of this model are 1) the cost of taking action and 2) the loss encountered 
if no protective action is taken but the event occurs. The cost-loss ratio characterises a user in the sense that it reflects a specific user-defined application. 
Building on this model, the so-called potential economic value is derived from the elements of the contingency table \citep{richardson2000,Richardson2011}.

When considering extreme events, we can hypothesise that their monetary impact is generally larger than the cost of preventive action.  
In this context, we can recall that, by design, the diagonal score (the score consistent with CPF) 
links the cost-loss ratio of a user with the rarity of the corresponding event: the rarer the event, the smaller the cost-loss ratio. 
More precisely, the CPF is designed as an optimal forecast for a user whose cost-lost ratio varies linearly with the event base rate.
This relationship is purely theoretical but helped derive a score with the useful meta-property of equitability \citep[see][for more details]{zbb2017}.

\begin{figure*}[hb]
\begin{center}  
  \includegraphics[width=0.95\textwidth]{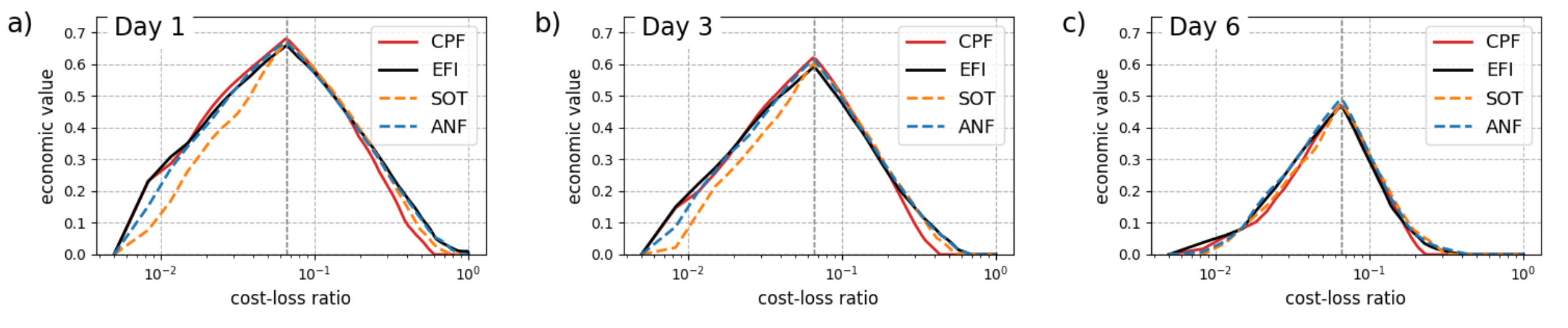}
\end{center}  
\caption{
    Potential economic value of daily precipitation forecasts as a function of a user's cost-loss ratio. Results at (a) day 1, (b) day 3, and (c) day 6 for Europe over Summer 2021. 
    Note that the x-axis is distorted (using a log-scale). 
}
\label{fig:pev}
\end{figure*}

Fig.~\ref{fig:pev} shows the potential economic value of CPF, EFI, SOT, and ANC as a function of the user's cost-loss ratio. 
Here again, we use 500 decision thresholds and results are drawn from the same contingency tables as for the plots in Fig.~\ref{fig:pdf} 
A log scale is used on the X-axis  
to emphasize forecast performance for users with a small cost-loss ratio. The user with a cost-loss ratio equal to the event base rate is 
indicated with a vertical line. By construction, the maximum forecast value is always reached precisely for this cost-loss ratio. 

CPF has a similar value as EFI for most users.
Users that should prefer EFI to CPF for decision-making are users with large cost-loss ratios, 
that is users with a potential loss close to the cost for preventive action. This type of user would take preventive action only 
when the event likelihood is high (the uncertainty is low). EFI appears more appropriate than CPF in that case.  
For situations where the user's cost-loss is close to the event base rate, CPF has more value than EFI, SOT, or ANC at day 1 and day 3.
With larger score uncertainty at longer time ranges, day 6 results show that all forecasts have an estimated economic value in the same ballpark at this time range.

\subsection{CPF reliability}
\label{sec:relia}
So far, CPF has been interpreted and assessed as an index, that is with no statistical meaning attached to its values. 
However, CPF is an ensemble summary statistic which is also interpretable as a probabilistic forecast.
So we propose to assess the CPF reliability because reliability is, along discrimination, a key attribute of a probabilistic forecast.
An example of what reliability means for CPF is the following: for all cases where CPF takes value 0.95, it is expected that the observations exceed the 95\% quantile 
of the climatology in 5\% of the time. This attribute can be checked with the help of a so-called reliability diagram.

\begin{figure*}
\begin{center}
  \includegraphics[width=0.95\textwidth]{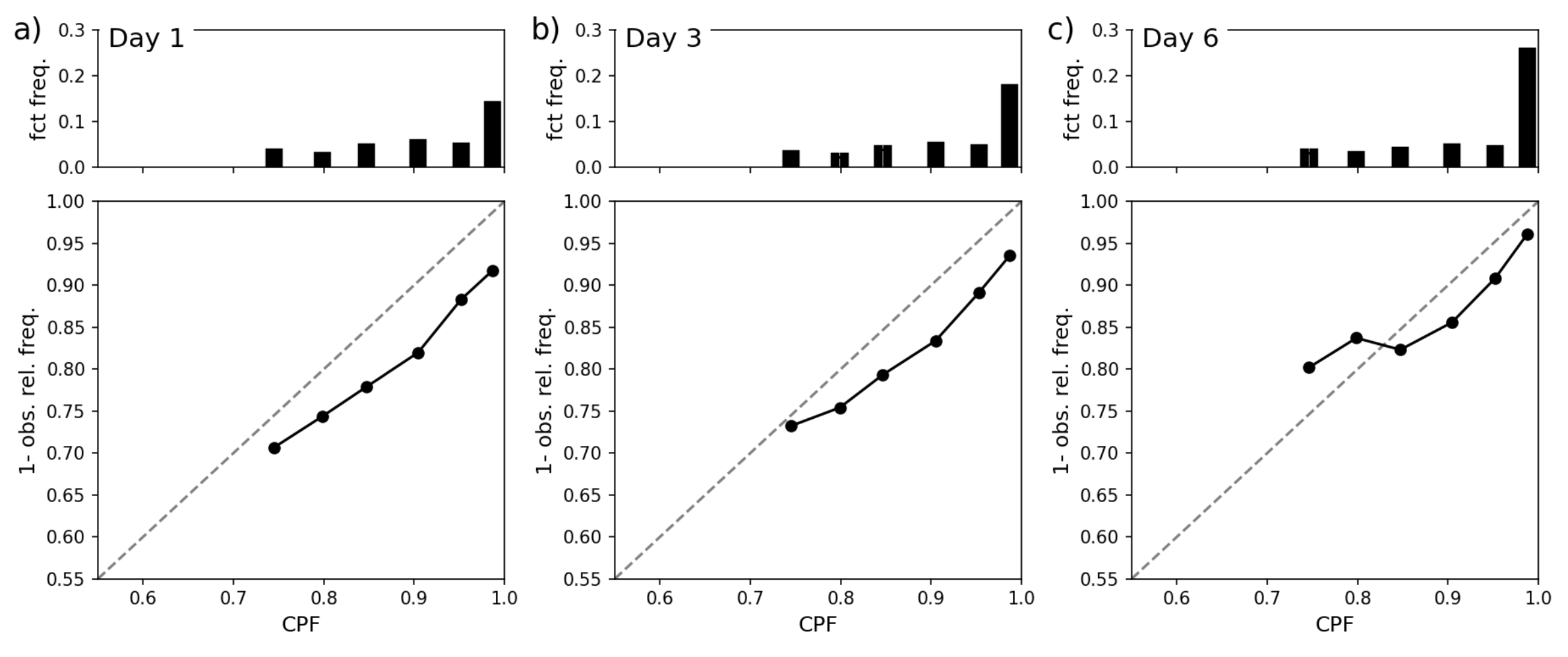}
\end{center}
\caption{ 
    Reliability diagram for CPF of daily precipitation. Results are shown for forecasts at lead times (a) day 1, (b) day 3, and (c) day 7. 
    In the lower panel, the observation relative frequency is aggregated for different CPF categories and
    perfect reliability is indicated with a dashed line. The upper panel shows the distribution of CPF in each category.
}
\label{fig:reliab}
\end{figure*}

The reliability diagrams for CPF of daily precipitation are shown in Fig.~\ref{fig:reliab}. In these plots, we show only cases where CPF is greater than 0.725, 
for categories centred around 0.75, 0.8, 0.85, 0.9, 0.95, and 0.99.
The bottom panels show the reliability curves that ideally should be close to the diagonal (dashed line).
The upper panel shows the percentage of cases when CPF falls within each of these categories. 

In terms of consistency with the observations, precipitation CPFs exhibit better reliability at longer lead 
times than at shorter ones. The ensemble forecast is notably under-spread over the first days of the forecast (day 1 and 3). 
This characteristic is reflected in the CFP reliability plot in the lower panel of Figs~\ref{fig:reliab}(a,b).
In the upper panel, we see that high values of CPF are more frequent at longer lead times, as already discussed in Section~\ref{sec:evol}. 

Reliability is an important forecast attribute for users who take forecasts at face value. 
As the CPF has an interpretation in probabilistic terms, forecast reliability ensures 
that the attached meaning to a forecast value and its actual meaning are aligned. 
Reliability is a necessary condition for optimal decision making \citep{wilks2018}.

\section{Actionable forecasts for seamless predictions of extreme events}
\label{sec:discuss}

An actionable forecast is defined as a single value forecast\footnote{sometime also referred to as a \textit{point forecast}, by opposition to an ensemble forecast.} discretised with a set of decision thresholds such that
a user can decide to take action when a threshold is exceeded by the forecast. 
Seamless forecasting across lead times would rely on actionable forecasts that could be used across time ranges, 
without the need to change the forecast definition or the set of decision thresholds 
with the forecast lead time. 

Potential discrimination was examined in the previous section. Now, we assess
the forecast information content communicated to the end user for decision-making. 
Forecast visualisation plays a key role in translating potential into actual discrimination ability.
The success of this translation is measured using a verification setting consistent with the forecast visualisation: 
the same set of decision thresholds is used for both applications. The rationale is that a piece of information that 
exists but is not communicated is not useful and has no value. This is typically the case when 
small probability values are attached to high-impact weather events, especially in the medium-range.

\subsection{Performance of actionable forecasts}

The forecasts are now discretised using a small set of decision thresholds that can be used for plotting, 
communication, and eventually decision-making purposes. For consistency, performance assessment relies on the same set of thresholds to build contingency tables.
The CPF decision thresholds are 0.85, 0.95, 0.98, 0.99, and 0.999, and the EFI ones are 0.3, 0.5, 0.6, 0.7, 0.8, and 0.9, 
and can be visualised in Figs~\ref{fig:flood} and \ref{fig:london}.
In these illustrations, the first decision threshold is represented 
by a contour line while the colour coding is used to show the other decision thresholds on the map. 
Note that the EFI visualisation in this study is identical to the one used on the ECMWF official web charts.
For SOT, the thresholds used for plotting SOT contours on the same charts are 0, 1, 2, 5, and 8. 
ANF is not included in this analysis because this forecast type does not belong to the current ECMWF portfolio for medium-range weather forecasts.

\begin{figure*}
\begin{center}
  \includegraphics[width=0.95\textwidth]{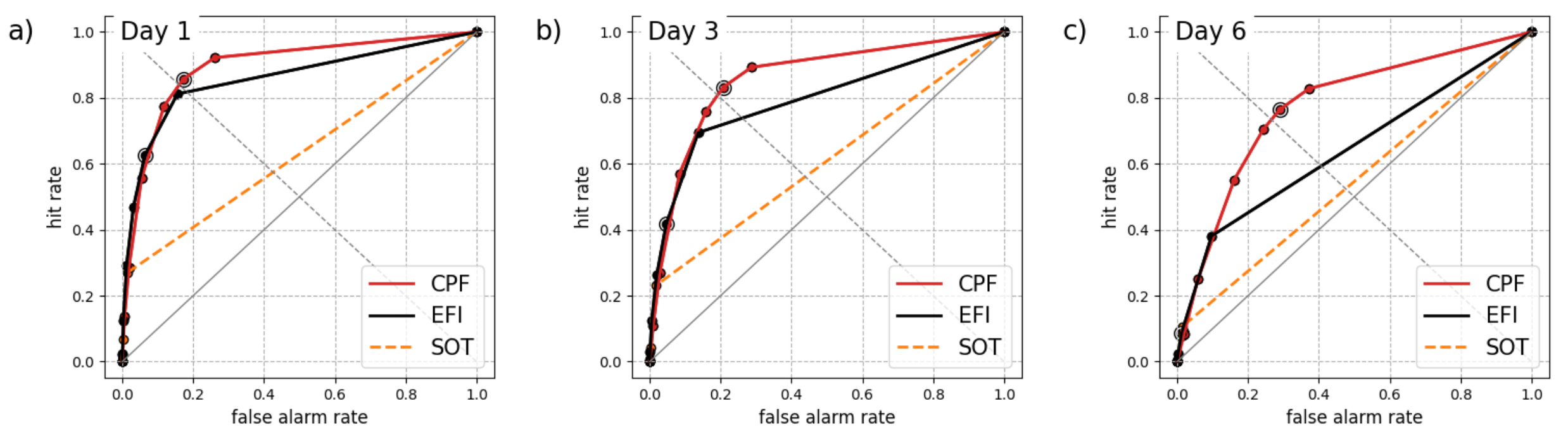}
\end{center}
    \caption{
    Discrimination ability of the actionable forecasts as plotted on ECMWF charts.
    Same as Fig. \ref{fig:pdf}  but when considering a fixed set of decision thresholds for each type of forecast. 
}
\label{fig:roc}
\end{figure*}

ROC curves built using these sets of decision thresholds are plotted in Fig.~\ref{fig:roc}.  Results are shown for
 CPF and EFI at 3 different lead times, day 1, day 3, and day 6. For each curve, one point is highlighted.  
This point corresponds to the combination ($H$,$F$) for a given decision threshold: the decision thresholds selected 
for illustration are 0.95 for CPF and 0.7 for EFI. In these plots, the ROC curves for SOT are also displayed for reference.

Two key messages emerge from the inspection of Fig.~\ref{fig:roc}. 
First, the empirical ROC curves are ``incomplete'' for EFI (and SOT) while they are almost complete for CPF. 
Second, the general performance associated with the selected decision thresholds (indicated with \( \circledbullet \)) varies with the forecast lead time for EFI but not for CPF.
These 2 aspects are closely related and contribute to CPF being an appropriate actionable forecast for high-impact weather, in particular at longer lead times. 

The EFI curves are not complete because the set of decision thresholds does not include values smaller than 0.5, but 
at day 6 for example, most of the EFI values are below 0.5 as illustrated in Fig.~\ref{fig:distri}(b).
By changing the EFI discretisation and adding smaller decision thresholds, more points would contribute  
to the ROC curve leading to a higher AUC. A finer discretisation of low EFI values would be particularly beneficial 
for forecasts at longer lead times while the current discretisation seems more appropriate for forecasts at shorter lead times. 

So, for EFI, an ideal set of decision thresholds would vary with forecast lead time.
To improve the EFI actual discrimination power, one would recommend a finer discretisation of low EFI values at longer lead times,
when the uncertainty is large and EFI generally small. By contrast, with CPF, a consistent set of decision thresholds can be used seamlessly 
throughout all lead times.

\subsection{Seamless prediction of extreme events}

Seamless forecasting is possible with CPF because the set of decision thresholds maintains their performance characteristics and meaning with lead time.

The highlighted points on the CPF curves (points associated with a given CPF decision threshold) do not dramatically change position on the curves with the forecast lead time.
A direct link exists between the position on a ROC curve and the optimal decision threshold for a given application. 
Indeed, a given combination of false alarm rate $F$ and hit rate $H$ is more appropriate for some users than for others.
In other words, for an application with a given cost-loss ratio, a forecaster can focus on a single decision threshold when using CPF. 
By contrast, the optimal decision threshold would vary with forecast lead time when using EFI (or SOT).

\begin{figure*}[t]
  \begin{center}
    \includegraphics[width=0.95\textwidth]{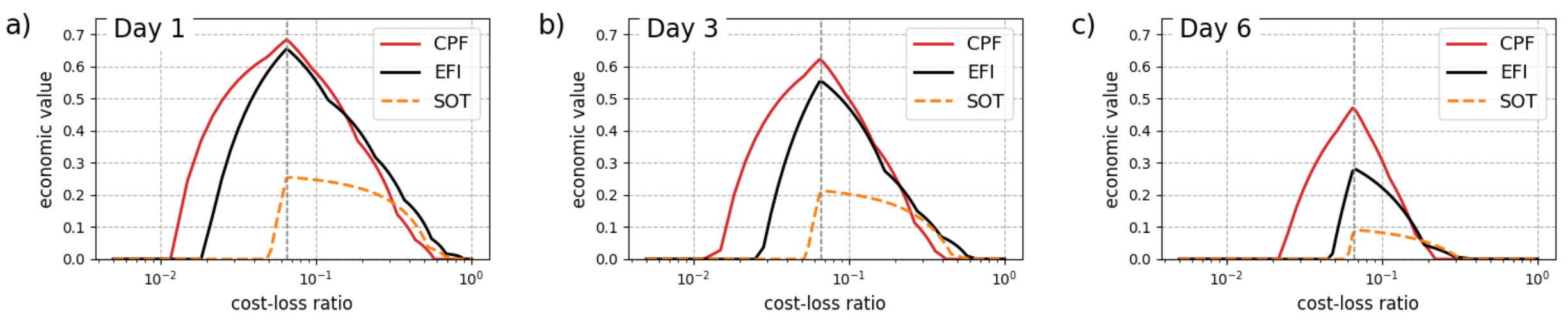}
  \end{center}
  \caption{
    Economic value of daily precipitation actionable forecast. 
    Same as Fig.~\ref{fig:pev} but when considering the same set of decision-threolds as in Fig.~\ref{fig:roc}, 
    that is the thresolds used for the forecast visualisation in Figs~\ref{fig:flood} and \ref{fig:london}. 
}
\label{fig:pev2}
\end{figure*}

The optimal decision threshold is known \textit{apriori} for CPF when CPF is well calibrated\footnote{at longer lead times, CPF is not far from calibration as illustrated in Fig.~\ref{fig:reliab}(c)}.
For example, a user with a cost-loss ratio of 0.05 should use the CPF with a decision threshold of 0.95.
More generally, a more complete ROC curve means more users would benefit from the forecast, especially users with a small cost-loss ratio. 
This is reflected in the economic value plots in 
Fig.~\ref{fig:pev2}. Here, the economic value of CPF, EFI, and SOT is estimated with the same contingency tables as for the ROC curves in Fig.~\ref{fig:roc}.

\begin{figure*}[t]
\vspace{0.8cm}
\begin{center}    
\includegraphics[width=0.49\textwidth,trim={0 0 0 0cm},clip]{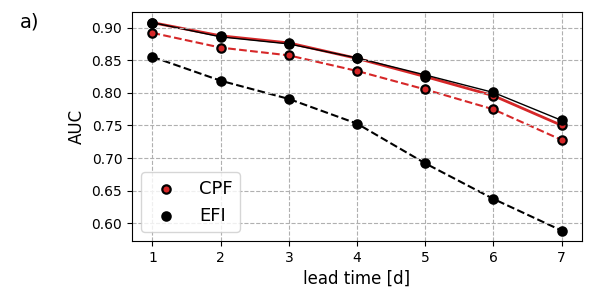}
\includegraphics[width=0.49\textwidth,trim={0 0 0 0cm},clip]{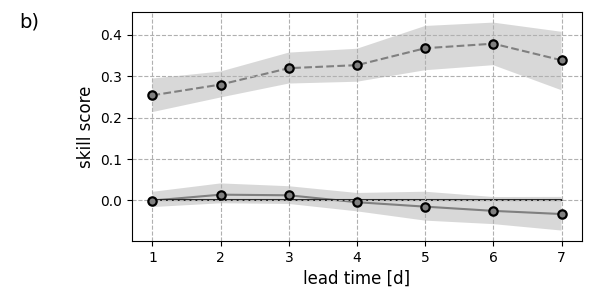}
\end{center}
\caption{
    (a) Area under the ROC curve as a function of the forecast lead time and 
    (b) corresponding skill score of CPF using EFI as a reference forecast. 
    The solid lines indicate the potential discrimination ability while the dashed lines indicate the actual discrimination ability 
    when considering a limited set of decision thresholds. 
    In (b), the shade represents the 5\% and 95\% confidence intervals as estimated with 5-day block boot-strapping.
    Forecast discretisation has a negative impact on EFI discrimination while CPF discrimination is preserved at all lead times.
}
\label{fig:auc}
\end{figure*}

Differences between potential and actual discrimination are visible when comparing ROC curves in Fig.~\ref{fig:pdf} and  Fig.~\ref{fig:roc}, respectively. 
We can also directly compare these differences by computing the area under the ROC curve (AUC). 
Relative differences in discrimination ability are assessed by computing the 
corresponding skill score measuring here the relative performance of CPF with respect to EFI:
\begin{equation}
    \frac{\text{AUC}_{CPF}-\text{AUC}_{EFI}}{1-\text{AUC}_{EFI}}.
\end{equation}
AUC results as a function of the forecast lead time are presented in Fig.~\ref{fig:auc}. 

In Fig.~\ref{fig:auc}(a), a large difference between potential discrimination (solid lines) and actual discrimination (dashed line) is 
an indication that the current way of communicating a forecast could be improved. A clear gap exists for EFI and this gap increases with lead time. 
As already noted by \cite{raynaud2018}, "[...] a more accurate calibration of optimal thresholds would be necessary in order to improve the EFI utilisation."
The difference between potential and actual discrimination is much smaller for CPF and is constant with lead time.

In Fig.~\ref{fig:auc}(b), CPF appears significantly better at actually discriminating than EFI at all lead times. 
The superiority of CPF in terms of actual discrimination increases with the forecast 
lead time reaching a skill score of almost 40\% at day 6. With CPF, a forecaster is able to actually discriminate the occurence of extreme precipitation events 
7 days ahead with the same ability as 4 days ahead when using EFI.

\section{Conclusion}
\label{sec:conc}

This study presents the singularity and performance of a new index for the early detection of high-impact weather events.
Case studies and results focus on daily precipitation, but the concept can be applied to any other weather variable. 
We show that :
\begin{enumerate}
\item the new index has a statistical meaning related to the concept of return-period and is directly linked to a proper score, 
\item the new index has similar \textit{potential} discrimination ability but a different evolution with lead time compared to all the other key indicators of extreme events 
 investigated here,
\item the new index can display a strong signal at longer lead times pointing to a potential high-impact weather event
leading to an enhanced \textit{actual} discrimination power on visualised maps,
\end{enumerate}

The new proposed index is derived from the crossing-point forecast (CPF) which is a consistent forecast with the diagonal score used 
for assessing ensemble forecasts with a focus on extreme events. The CPF can be interpreted as a "worst-case scenario in a probabilistic sense" 
for which the concept of reliability holds and can be further assessed. CPF encompasses a duality risk/intensity in its interpretation: 
its value refers at the same time to a risk level and to an event intensity
that can be expressed in terms of a return period. Conveniently, the index takes a value between 0 and 1, so it can be used and visualised as an index. 
But more importantly, its (statistical) meaning and characteristics do not vary with lead time.

A comparison with the well-established extreme forecast index EFI reveals that the new index has a similar potential discrimination ability of extreme precipitation events
and a lower correlation with EFI than the shift of tail SOT or the ensemble mean standardised anomaly ANC.
The new index is the only forecast among those analysed here that exhibits a stronger signal for extreme events at longer rather than shorter lead times.
This new type of forecast is particularly well-suited for users sensitive 
to missed events but could accept a higher level of false alarms.
Typically, the users who should most benefit from the new index have a small cost-loss ratio, in other words,
their cost of taking a preventive action is rather lower compared to the potential loss in case of a missed event.
The new index is also better at distinguishing the most extreme events from lower-intensity precipitation events. 

Close to the event, CPF and EFI converge to similar ``solutions'', but at longer lead times, we observe a complementarity between CPF and EFI: 
EFI tends to be muted which reduces false alarms while CPF tends to have a large-scale signal which reduces missed events.
In most cases, EFI does not exceed 0.5 at day 6 because it is highly correlated with the ensemble mean anomaly and
dampened as the ensemble spread increases. While it might be difficult to trigger forecasters' attention on an event with a small probability of occurrence in the forecast 
or an EFI close to 0, with CPF, low-probability events are put forward when the forecast risk exceeds the climatological risk. 
As a result, a signal for a rare event can be present even when the forecast uncertainty is large,
 but with the caveat that this signal requires confirmation with the following forecast runs to limit false alarms.
As illustrated with two case studies, precipitation events with low probability and potentially high impact can be seamlessly captured and 
communicated at the medium-range time scale with the help of the new index.



\begin{thebibliography}{27}
  \providecommand{\natexlab}[1]{#1}
  \providecommand{\url}[1]{\texttt{#1}}
  \providecommand{\urlprefix}{URL }
  \expandafter\ifx\csname urlstyle\endcsname\relax
    \providecommand{\doi}[1]{doi:\discretionary{}{}{}#1}\else
    \providecommand{\doi}{doi:\discretionary{}{}{}\begingroup
    \urlstyle{rm}\Url}\fi
  
  \bibitem[{{Ben Bouall\`egue}(2021)}]{zbb2021}
  {Ben Bouall\`egue} Z. 2021. On the verification of the crossing-point forecast.
    \emph{Tellus A: Dynamic Meteorology and Oceanography} \textbf{73}(1): 1--10,
    \doi{10.1080/16000870.2021.1913007}.
  
  \bibitem[{{Ben Bouall\`egue} \emph{et~al.}(2023){Ben Bouall\`egue}, Clare,
    Magnusson, Gascon, Maier-Gerber, Janousek, Rodwell, Pinault, Dramsch, Lang,
    Raoult, Rabier, Chevallier, Sandu, Dueben, Chantry and
    Pappenberger}]{zbb2023bams}
  {Ben Bouall\`egue} Z, Clare MCA, Magnusson L, Gascon E, Maier-Gerber M,
    Janousek M, Rodwell M, Pinault F, Dramsch JS, Lang STK, Raoult B, Rabier F,
    Chevallier M, Sandu I, Dueben P, Chantry M, Pappenberger F. 2023. The rise of
    data-driven weather forecasting. \emph{arXiv}
    \doi{10.48550/arXiv.2307.10128}.
  
  \bibitem[{{Ben Bouall\`egue} \emph{et~al.}(2018){Ben Bouall\`egue}, Haiden and
    Richardson}]{zbb2017}
  {Ben Bouall\`egue} Z, Haiden T, Richardson DS. 2018. {The diagonal score:
    definition, properties, and interpretations}. \emph{Quart. J. Roy. Meteor.
    Soc.} \textbf{144}(714): 1463--1473, \doi{10.1002/qj.3293}.
  
  \bibitem[{{Ben Bouall\`egue} \emph{et~al.}(2019){Ben Bouall\`egue}, Magnusson,
    Haiden and Richardson}]{zbb19}
  {Ben Bouall\`egue} Z, Magnusson L, Haiden T, Richardson DS. 2019. Monitoring
    trends in ensemble forecast performance focusing on surface variables and
    high-impact events. \emph{Quart. J. Roy. Meteor. Soc.} \textbf{145}(721):
    1741--1755, \doi{10.1002/qj.3523}.
  
  \bibitem[{{Ben Bouallègue} and Richardson(2022)}]{zbb2022}
  {Ben Bouallègue} Z, Richardson DS. 2022. On the roc area of ensemble forecasts
    for rare events. \emph{Weather and Forecasting} \textbf{37}(5): 787 -- 796,
    \doi{10.1175/WAF-D-21-0195.1}.
  
  \bibitem[{Boisserie \emph{et~al.}(2016)Boisserie, Descamps and
    Arbogast}]{boisserie2016}
  Boisserie M, Descamps L, Arbogast P. 2016. Calibrated forecasts of extreme
    windstorms using the extreme forecast index (efi) and shift of tails (sot).
    \emph{Weather and Forecasting} \textbf{31}(5), \doi{10.1175/WAF-D-15-0027.1}.
  
  \bibitem[{Dutra \emph{et~al.}(2013)Dutra, Diamantakis, Tsonevsky, Zsoter,
    Wetterhall, Stockdale, Richardson and Pappenberger}]{dutra2013}
  Dutra E, Diamantakis M, Tsonevsky I, Zsoter E, Wetterhall F, Stockdale T,
    Richardson D, Pappenberger F. 2013. The extreme forecast index at the
    seasonal scale. \emph{Atmospheric Science Letters} \textbf{14}(4): 256--262,
    \doi{10.1002/asl2.448}.
  
  \bibitem[{Fundel \emph{et~al.}(2010)Fundel, Walser, Liniger, Frei and
    Appenzeller}]{fundel2010}
  Fundel F, Walser A, Liniger MA, Frei C, Appenzeller C. 2010. Calibrated
    precipitation forecasts for a limited-area ensemble forecast system using
    reforecasts. \emph{Monthly Weather Review} \textbf{138}(1),
    \doi{10.1175/2009MWR2977.1}.
  
  \bibitem[{Fundel \emph{et~al.}(2019)Fundel, Fleischhut, Herzog, Göber and
    Hagedorn}]{fundel2019}
  Fundel VJ, Fleischhut N, Herzog SM, Göber M, Hagedorn R. 2019. Promoting the
    use of probabilistic weather forecasts through a dialogue between scientists,
    developers and end-users. \emph{Quarterly Journal of the Royal Meteorological
    Society} \textbf{145}(S1): 210--231, \doi{10.1002/qj.3482}.
  
  \bibitem[{Guan and Zhu(2017)}]{guan2017}
  Guan H, Zhu Y. 2017. Development of verification methodology for extreme
    weather forecasts. \emph{Weather and Forecasting} \textbf{32}(2): 479 -- 491,
    \doi{10.1175/WAF-D-16-0123.1}.
  
  \bibitem[{Haiden \emph{et~al.}(2021)Haiden, Janousek, Vitart, {Ben Boua\'egue},
    Ferranti and Prates}]{haiden2021}
  Haiden T, Janousek M, Vitart F, {Ben Boua\'egue} Z, Ferranti L, Prates F. 2021.
    {Evaluation of ECMWF forecasts, includuing 2021 upgrade}. \emph{ECMWF
    Technical Memorandum} \textbf{884}.
  
  \bibitem[{Lalaurette(2003)}]{lalaurette2003}
  Lalaurette F. 2003. Early detection of abnormal weather conditions using a
    probabilistic extreme forecast index. \emph{Quart. J. Roy. Meteor. Soc.}
    \textbf{129}: 3037--3057, \doi{doi: 10.1256/qj.02.152}.
  
  \bibitem[{Lavers \emph{et~al.}(2016)Lavers, Pappenberger, Richardson and
    Zsoter}]{lavers2016}
  Lavers DA, Pappenberger F, Richardson DS, Zsoter E. 2016. {ECMWF Extreme
    Forecast Index} for water vapor transport: A forecast tool for atmospheric
    rivers and extreme precipitation. \emph{Geophysical Research Letters}
    \textbf{43}(22): 11,852--11,858, \doi{10.1002/2016GL071320}.
  
  \bibitem[{Lavers \emph{et~al.}(2018)Lavers, Richardson, Ramos, Zsoter,
    Pappenberger and Trigo}]{lavers2018}
  Lavers DA, Richardson DS, Ramos AM, Zsoter E, Pappenberger F, Trigo RM. 2018.
    Earlier awareness of extreme winter precipitation across the western {Iberian
    Peninsula}. \emph{Meteorological Applications} \textbf{25}(4): 622--628,
    \doi{10.1002/met.1727}.
  
  \bibitem[{Lavers \emph{et~al.}(2017)Lavers, Zsoter, Richardson and
    Pappenberger}]{lavers2017}
  Lavers DA, Zsoter E, Richardson DS, Pappenberger F. 2017. An assessment of the
    {ECMWF Extreme Forecast Index} for water vapor transport during boreal
    winter. \emph{Weather and Forecasting} \textbf{32}(4): 1667 -- 1674,
    \doi{10.1175/WAF-D-17-0073.1}.
  
  \bibitem[{Leutbecher(2019)}]{leutbecher2018}
  Leutbecher M. 2019. Ensemble size: How suboptimal is less than infinity?
    \emph{Q. J. R. Meteorolog. Soc.} \textbf{145}(Suppl. 1): 107--128,
    \doi{10.1002/qj.3387}.
  
  \bibitem[{Murphy and Winkler(1987)}]{murphy87}
  Murphy AH, Winkler RL. 1987. A general framework for forecast verification.
    \emph{Mon. Wea. Rev.} \textbf{115}: 1330--1338.
  
  \bibitem[{Neal \emph{et~al.}(2014)Neal, Boyle, Grahame, Mylne and
    Sharpe}]{neal2014}
  Neal RA, Boyle P, Grahame N, Mylne K, Sharpe M. 2014. Ensemble based first
    guess support towards a risk-based severe weather warning service.
    \emph{Meteorological Applications} \textbf{21}(3): 563--577,
    \doi{10.1002/met.1377}.
  
  \bibitem[{Petroliagis and Pinson(2014)}]{petroliagis2014}
  Petroliagis TI, Pinson P. 2014. Early warnings of extreme winds using the
    {ECMWF Extreme Forecast Index}. \emph{Meteorological Applications}
    \textbf{21}(2): 171--185, \doi{10.1002/met.1339}.
  
  \bibitem[{Prates and Buizza(2011)}]{prates2011}
  Prates F, Buizza R. 2011. {PRET}, the {Probability of RETurn}: a new
    probabilistic product based on generalized extreme-value theory.
    \emph{Quarterly Journal of the Royal Meteorological Society}
    \textbf{137}(655): 521--537, \doi{10.1002/qj.759}.
  
  \bibitem[{Raynaud \emph{et~al.}(2018)Raynaud, Touzé and
    Arbogast}]{raynaud2018}
  Raynaud L, Touzé B, Arbogast P. 2018. Detection of severe weather events in a
    high-resolution ensemble prediction system using the extreme forecast index
    ({EFI}) and shift of tails ({SOT}). \emph{Weather and Forecasting}
    \textbf{33}(4), \doi{10.1175/WAF-D-17-0183.1}.
  
  \bibitem[{Richardson(2000)}]{richardson2000}
  Richardson DS. 2000. Skill and relative economic value of the {ECMWF} ensemble
    prediction system. \emph{Q. J. R. Meteorol. Soc.} \textbf{126}: 649--667,
    \doi{10.1002/qj.49712656313}.
  
  \bibitem[{Richardson(2011)}]{Richardson2011}
  Richardson DS. 2011. Economic value and skill. In: \emph{Forecast Verification:
    A Practitioner's Guide in Atmospheric Science}, Jolliffe IT, Stephenson DB\
    (eds), John Wiley and Sons, pp. 167--184.
  
  \bibitem[{Tempest \emph{et~al.}(2023)Tempest, Craig and Brehmer}]{tempest2023}
  Tempest KI, Craig GC, Brehmer JR. 2023. Convergence of forecast distributions
    in a 100,000-member idealised convective-scale ensemble. \emph{Quarterly
    Journal of the Royal Meteorological Society} \textbf{149}(752): 677--702,
    \doi{10.1002/qj.4410}.
  
  \bibitem[{Tsonevsky \emph{et~al.}(2018)Tsonevsky, Doswell and
    Brooks}]{tsonevsky2018}
  Tsonevsky I, Doswell CA, Brooks HE. 2018. Early warnings of severe convection
    using the {ECMWF Extreme Forecast Index}. \emph{Weather and Forecasting}
    \textbf{33}(3): 857 -- 871, \doi{10.1175/WAF-D-18-0030.1}.
  
  \bibitem[{Wilks(2018)}]{wilks2018}
  Wilks D. 2018. Enforcing calibration in ensemble postprocessing. \emph{Quart.
    J. Roy. Meteor. Soc.} \textbf{144}(710): 76--84, \doi{10.1002/qj.3185}.
  
  \bibitem[{Zsoter(2006)}]{zsoter2006}
  Zsoter E. 2006. Recent developments in extreme weather forecasting. \emph{ECMWF
    Newsletter} \textbf{107}: 8--17.
  
  \end{thebibliography}
\end{document}